\documentclass{article}

\usepackage{graphicx}

\usepackage{epsf}

\begin{document}

\title{Tilt Angles of Solar Filaments over the Period 1919-2014}
\author{Tlatov~A. G., Kuzanyan~ K.M. and  Vasil'yeva~V.V }


\maketitle

\begin{abstract}
The spatial and temporal distributions of solar filaments were analyzed using data from the Meudon Observatory for the period 1919-2003 and the  Kislovodsk Mountain Astronomical Station for the period 1979-2014.
We scanned $H_\alpha$ solar synoptic charts on which the filaments were isolated and digitized. The data on each filament comprise its location, length, area, and other geometrical characteristics.
The temporal distributions of the number and total length of the filaments have been obtained. We also found latitudinal migration of filament locations  with the solar cycle, and analyzed the longitudinal distribution and asymmetry of filaments in the northern and southern hemispheres, and other properties of their distribution.

The tilt angles of filaments with respect the solar equator ($\tau$) were analyzed. On average, the eastern tips of filaments are closer to the poles than the western ones ($\tau \sim 10^\circ$). On the other hand, the filaments in the polar regions ( $\theta>50^\circ$, where $\theta$ is the latitude)
 usually have negative tilts ($\tau <0^\circ$). The tilt angles vary with the phases of the 11 year sunspot cycle and are at their highest values in the epoch of the activity maximum. In the century-long modulation of the solar activity (Gleissberg cycle), the mean tilt angles of filaments in the mid-latitude zone ($\theta \sim \pm 40^\circ$) were maximum in the middle of the 20th century in solar sunspot cycles 18-19.
We hereby propose using the statistical properties of solar filaments as an additional coherent measure of manifestation of the solar cycle which covers all latitudes and for which almost a century long systematically calibrated data series is available.

\end{abstract}



\section{Introduction}
     \label{S-Introduction}

The appearance of filaments on the solar disk is associated with the distribution of large-scale magnetic fields and magnetic fields of active regions (AR). The filaments are observed virtually at all latitudes in the solar atmosphere.

Systematic $H_\alpha$ observations have been carried out approximately since 1915 at the Kodaikanal Observatory (India), since 1919 at the Meudon Observatory (France) (spectroscopic observations), and since 1959 at the Kislovodsk Mountain Astronomical Station of the Main (Pulkovo) Astronomical Observatory of Russian Academy of Sciences. The filaments observed in the $H_\alpha$ line in the solar chromospheres are one of the basic indices of solar activity (see the review by \cite{Kiepenheuer}). Solar filaments arise in the vicinity of the magnetic field polarity inversion line (\cite{McIntosh}, \cite{Makarov83},  \cite{Karachik}). As a rule, they are divided into several types (\cite{D'Azambuja23}, \cite{D'Azambuja28},  \cite{D'Azambuja48}):  AR filaments, quiescent filaments, and polar filaments (\cite{Mouradian94}). The former are usually short (about 10 Mm), have small lifetimes (hours to days), and are located quite low ($<10$ Mm). The filaments outside the active regions may be hundreds of megameters long, have lifetimes up to a few rotations, and are located at a height of about 100 Mm (\cite{Mackay10}).
While the basic properties of solar filaments have long been known, filament channels and filaments have been classified more recently in terms of their chirality (\cite{Martin94}). They classified filament channels as either  dextra  or inistral depending on the
direction of the axial component of the field as seen by an observer standing on the positive-polarity side of the channel.

With respect to the polarity inversion line, the filaments are also divided into several groups. E. g., they can be located within one or several bipolar active regions (\cite{Tang}) or between diffuse magnetic fields of different polarity (\cite{Mackay08}).

Despite a century of observations and investigation of the solar filaments the physical mechanisms of their formation are not yet fully understood (see the review by \cite{Martin98}). However, the role of solar filaments in the occurrence of solar flares, coronal mass ejections, and geomagnetic disturbances is extremely high (\cite{Wang}).

The objective of this work is to study long-term variations in the filament characteristics by analyzing digitized $H_\alpha$ synoptic maps for the period 1919-2014. We have mainly focused on the change of the filament tilt angles relative to the heliographic coordinates.

\section{Original and digitized data} 
      \label{S-general}

We have used two types of the source data. The first type is Meudon synoptic maps. In the period 1919-1989, the Meudon Observatory published synoptic maps with plotted long-lived eruptive filaments, centers of activity, and faculae (\cite{Mouradian98}). The data are partly processed and are available in tabular form on the Web site (\textbf{$http://www.ngdc.noaa.gov/stp/space-weather/solar-data/solarfeatures/
prominences-filaments/filaments/$}). The tables contain information on the coordinates, time, and other characteristics of the filaments for two time intervals (1919-1956 and 1957-2000) with the addition of data from other observatories. Later, the procedure has been changed, and since 1990 the maps have been digitized and plotted.  They are provided in  digital format  (\cite{Mouradian98b}).

In this work, we have used the original synoptic maps. Thus, the maps for the period 1919-1989 (rotations 876-1850) have been scanned from atlas (\cite{Meudon48}). The maps in the new format for the synoptic rotations 1824-1850 (1989-1991) and rotations 1931-2008 (1998-2003) are available on the Web site ($http://bass2000.obspm.fr/lastsynmap.php$).

The second group of the source data comprises the maps of Kislovodsk Mountain Station for the period 1979-2014. They are based on daily chromospheric observations with the use of an $H_\alpha$ filter with the bandwidth of $0.25{\AA}$. The data are regularly processed, and daily maps of solar activity are provided. Then, the positions of the filaments are marked on synoptic charts.
Unlike the Meudon synoptic maps, the Kislovodsk ones give the polarity inversion lines of the large-scale magnetic field (\cite{Makarov83}). During 1979-1995, the $H_\alpha$ synoptic charts were published in the Solnechnye Dannye (Solar Data) Bulletin, and thereafter, it has been carried on in the electronic version of the Bulletin on the Web site $http://www.gao.spb.ru/english/database/sd/sinoptic{\_}charts.htm$. At present, the scanned maps are available for the synoptic rotations 1677-2155.

Besides the synoptic chart, there are tables that contain characteristics of solar filaments since 1959 till the present (\textbf{$http://en.solarstation.ru/archive$}).
For the  purposes like, for example, studies of some characteristics with the use of other types of observational data or determination of the filament geometric parameters, the filament images need to be vectorized. The Kislovodsk Mountain Astronomical Station provides data on solar filaments both in the form of daily images and $H_\alpha$ synoptic charts. Since 2004, the procedures of identification of filaments, determination of their characteristics, and plotting them on synoptic charts are automated (\cite{Tlatov}).
Other algorithms for automatic filament detection have been developed and implemented into various software and data products by e.g., (\cite{Bernasconi}; \cite{Yuan};  \cite{Hao}; \cite{Schuh}).

Figure~\ref{Fig1} shows an example of the Meudon and Kislovodsk synoptic charts. The original maps represent solar filaments schematically showing only their median line and provisional width. Special software was developed to digitize the filaments. The procedure is as follows. First, the synoptic chart boundaries and coordinate grid are applied to the scanned image. Then, the filaments are vectorized. In this process, the operator draws the median line along the filament and chooses an appropriate width in semi-automatic mode.

The following parameters were defined for each filament: mean coordinates of the filament and coordinates of its tips, length, latitudinal and longitudinal dimensions, tilt angle  and polarity of magnetic field. The total number of digitized filaments is 81013 from Meudon data for the period 1919-2003 and 17787 from Kislovodsk data.

The database comprises both the filaments associated with active region magnetic dipoles and quiescent filaments arising between diffuse magnetic configurations. Figure ~\ref{Fig2} represents schematically the positions and tilt angles of the filaments of various  types.

For further analysis, we constructed a combined series comprising the data from Meudon synoptic charts for the period 1919-1985 (rotations 876-1769) and from Kislovodsk synoptic charts for the period after 1985 (rotations 1770-2155). Altogether, the combined series included the data for the period 1919-2014 on 74572 filaments.

Figure~\ref{Fig3}a illustrates variations in the number of filaments on synoptic charts. One can clearly identify an 11-year mode, which reflects variation of solar activity. On the other hand, the average number of filaments has been increasing during 1919-2005. This may be due to the changes in the methods and quality of observations. A more adequate characteristic can be the  total length of filaments on the synoptic chart represented in Figure~\ref{Fig3}b. According to the Meudon data, the total length of filaments on the synoptic maps before 1988 has been more stable. This is probably due to the fact that the new methods adopted in 1988 allowed one seeing weak filaments and filament channels. The Kislovodsk and Meudon maps for cycles 21 and 22 appear more similar as concerned with the length of the filaments rather than their number, probably, because the filaments on Meudon maps are plotted as a chain of individual feature lines, while on Kislovodsk maps they are  represented by solid lines.

Figure~\ref{Fig4} represents the distribution of the lengths of filaments of the 1919-2014 composite series. The distribution of the relative number of filaments versus the logarithm of their length is close to the lognormal distribution with a peak at about 70 Mm.

\section{Analysis of data} 
  \label{S-labels}

Filaments can be efficiently used as an activity index when analyzing long-term variations of solar activity.
Solar filaments are present virtually everywhere from the equator to the highest latitudes.
Figure~\ref{Fig5} shows the latitude-time diagram of the centers of digitized solar filaments.
There are two groups of filaments: filaments associated with the polar drift of the large-scale magnetic field and those associated with the distribution of sunspots.
Polar filaments appear on the polarity inversion lines of the large-scale magnetic field in high latitudes. At the
minimum and the rise phases of sunspot activity cycle, the filament centers drift towards the poles, ceasing their motion in the epoch of the cycle maximum.
There have been several branches of the poleward drift in some solar cycles, e.g., cycles 18, 19, and 20 in the northern hemisphere and in cycles 18 and 21 in the southern hemisphere. In this case, one can see the so-called "triple reversal" of the large-scale magnetic field polarity (\cite{Makarov83}; \cite{Li}). The drift velocity depends on the height of the activity cycle (\cite{Makarov01}, see their Table 3 for details).

The other group of filaments is associated with sunspot activity. These filaments appear in the mid-latitudes and equatorial zones following the butterfly diagram of sunspot distribution. The number of the filaments is maximum at latitudes of about $30^\circ$
(Figure~\ref{Fig5})
which is somewhat higher than for the maximum of sunspot distribution. Contrary to the latitudinal distribution of sunspots, a lot of filaments appear in immediate proximity to the equator.

Likewise the number filament, their average length is getting higher in low and middle latitudes.
Although the total length of filaments decreases to high latitudes, their longitudinal extent increases with latitude
from the average of about $12^\circ$ near the equator to about $27^\circ$ near the pole.
The longitudinal extent of filaments in the southern hemisphere in the period 1919-2014 was somewhat longer than on the northern one.

\section{Tilt angles of filaments}

In the context of different aspects of the solar dynamo, an important characteristic is the angle ($\tau$)
of the north-south tilt of the active region magnetic dipoles.
It is believed that  the tilt angle for sunspots and bipolar groups
determines the efficiency of conversion of toroidal fields into the poloidal magnetic field (\cite{Tlatov14}). Solar filaments usually lie along the magnetic polarity inversion lines. Measurements of the tilt angles may help us understand the mechanism of generation of the solar magnetic field.
To find the tilt angle, we draw a straight line through each filament approximating it by the least square method, and measured its inclination. The filaments whose eastern tips are closer to the poles than the western ones were given positive tilts, and negative if vice versa.

Figure~\ref{Fig6} represents the relative distribution of the tilt angle $\tau$ of solar filaments calculated with respect to their length as a weighting factor.
The distribution is strongly asymmetric about $\tau=0^\circ$. Although most filaments have the tilt close to $0^\circ$, the mean tilt angle is positive.
The total number of filaments with negative tilts is 24316, which is about a half of the number of the filaments with positive tilt angles. 16700 filaments of the latter group arise in the  sunspot formation zone, i.e., at latitudes below $\pm40^\circ$. This is quite remarkable.

For the filaments which are located between the isolated
bipolar groups of plages and active regions, we would expect their tilt angles to be negative. For the filaments associated with sunspots, the tilt angles must be close to normal to the magnetic axis of the dipoles (see the schemes in (\cite{Tang}, \cite{Mackay08}).
It is likely that the filaments plotted on synoptic maps mainly originate at the boundaries of large-scale diffuse magnetic configurations arising in the process of the decay of active regions (\cite{Mackay08}).

 The mean tilt angles changes with the phase of the cycle (Figure~\ref{Fig7}) being the highest in the epoch of solar maximum. We may note that the smoothed tilt angle values also could  vary with the secular (Gleissberg) cycle attaining their maximum value in the middle of 20th century with the only exception of the present cycle 24.

Figure~\ref{Fig7} also shows the cycle-mean tilt angles  with activity cycles. It also shows the cycle-mean tilt angles of filaments averaged over the period from $T^{n}_{min}$  to $T^{n+1}_{min}$ for different cycles. The maximum tilt angle has been recorded in cycle 19. In the prolonged minimum solar activity of cycles 23/24 the number of observed filaments was really small (see Figure 5). This may be a possible reason why the average tilt turned to negative values then. This issue is worth further investigation. There is a positive correlation between the amplitude of the activity cycle and the tilts of filaments. The correlation coefficient of linear regression for cycles 15-23 is $r=0.84$.

The tilt angles of filaments vary with their length
 increasing from about $6^\circ$ to $16^\circ$ for relatively shorter filaments. The maximum positive tilt angles correspond to the length of $\sim400$~Mm. For further longer filaments the tilt angles are lower  while there are fewer of them and for long filaments our definition of tilt by approximation of a straight line may not be that robust.

 The latitudinal distributions of the relative number of filaments with positive $\tau>0$ and negative $\tau<0$ tilts differ from each other (Figure~\ref{Fig8}). The number of filaments with positive tilt angles is maximum at the latitudes of about $20-30^\circ$, while the distribution of filaments with negative tilts number peaks at the vicinity of the equator and at latitudes about $50^\circ$.

The tilt angles of filaments also depend on latitude. Figure~\ref{Fig9} illustrates the latitudinal distribution of the tilt angles separately for the filaments with negative and positive tilts. The absolute value of the tilt angle is maximum in the low-latitude zone and decreases with absolute latitude. Since the total number of filaments with positive tilt is larger than the number of those with negative tilts, the distribution demonstrates on average mainly positive tilts with exception for the polar regions (Figure~\ref{Fig9}c).

The increase of the tilt angles in absolute value for the filaments with $\tau<0$ (Figure~\ref{Fig9}b) can probably be explained by the Joy law for sunspot dipoles. The tilt angles of sunspots increase with latitude but noting that filaments are located on the polarity inversion lines, one may expect them to be normal to the magnetic axis of the dipoles (cf. other works on filaments: \cite{Martin94}; \cite{Pevtsov}).

For the filaments with positive tilts, the tilt angles have maximum of about $30^\circ$ at latitudes of about $20^\circ$ and decrease to the polar regions.
Figure~\ref{Fig10} represents the time-latitude distribution of the tilt angles. The red areas correspond to positive tilts and the blue ones to the negative tilts. In the mid-latitude zone, the tilt is mainly positive. At high latitudes, it is mainly negative, especially for the filaments associated with the polar drift (Figure~\ref{Fig5}).

\section{Active Region Filaments and Magnetic Fields}

In this work, we have analyzed the solar filaments recorded on the Meudon and Kislovodsk $H_\alpha$ synoptic charts for about 100 year long period.

We have found that different types of filaments as mentined in introduction  (Figure~\ref{Fig2}) 
have different characteristic tilt angles $\tau$. Variations in the tilt angles in different latitudinal zones  for low- and mid-latitude filaments (in the range $\theta<\pm40^\circ$) and polar filaments ($\theta>50^\circ$) in different latitudinal zones during cycles 15-23 are shown in Figure~\ref{Fig11}. The number of polar filaments is 10309 or about $12\%$ of the total number. The number of filaments with $\tau<0$ at mid latitudes ($\theta<40^\circ$) is 18303 or $\sim22\%$ of the total number. Low-latitude filaments have mainly positive tilt angles $\sim10-13^\circ$. Their maximum was attained in the middle of the 20th century. The tilt angles of polar filaments are usually smaller and were negative in cycles 15-20.

Most of the filaments are located in the mid-latitude zone. These are both the filament associated with active regions and those associated with diffuse magnetic fields. It may be suggested (\cite{Mackay08}) that the filaments of different types have the tilt angles of different sign. Figure~\ref{Fig12} illustrates the time variation of filaments with negative and positive tilt angles. The mean tilt angle of filaments with $\tau>0$ decreased during cycles 17-22. Filaments with positive tilts $\tau>0$ have a long-term trend with $8\%$ variability (Figure~\ref{Fig12}a).

\begin{figure}
    \centering
    \includegraphics[width=5.0in]{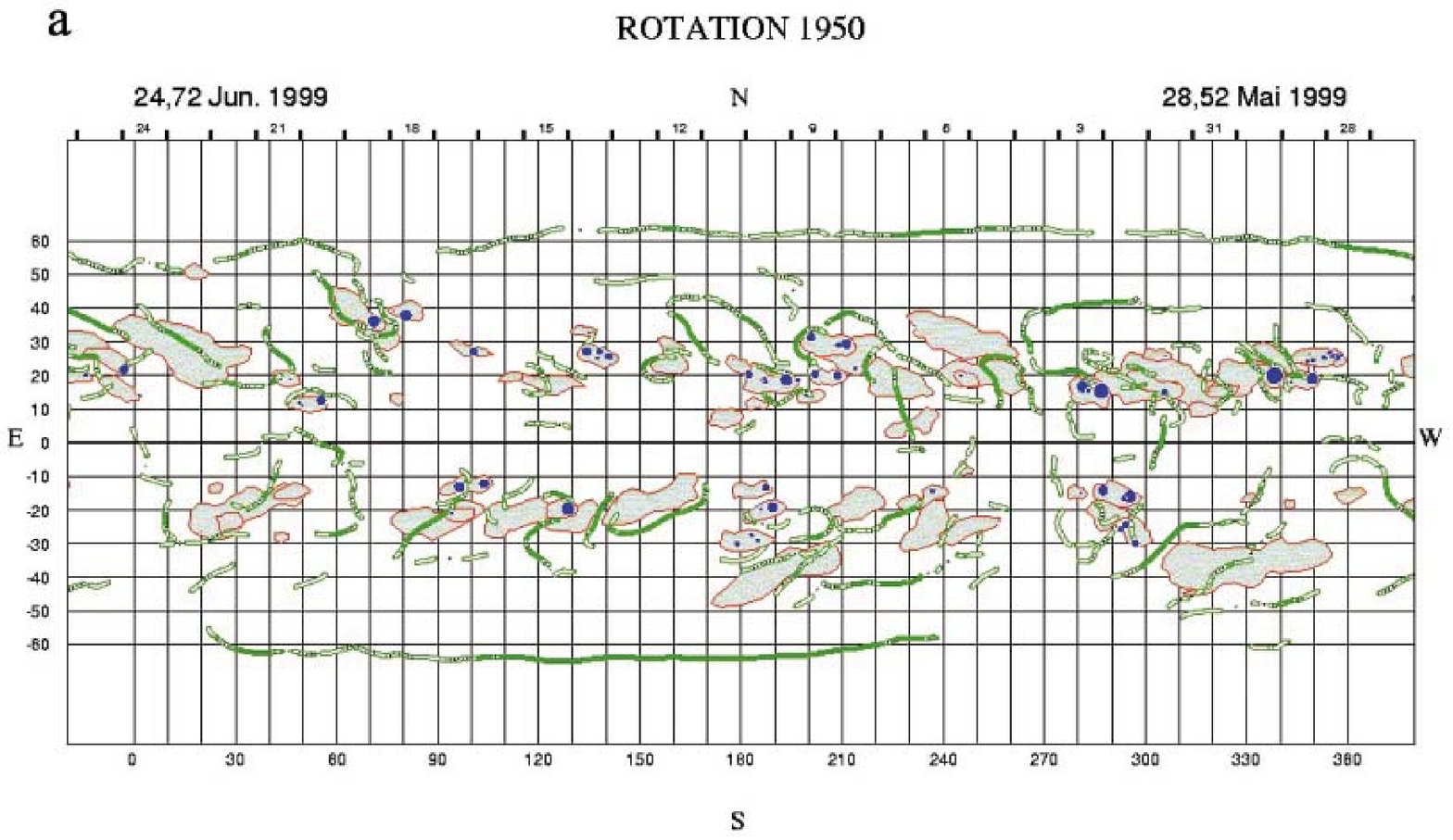}
   \includegraphics[width=5.0in]{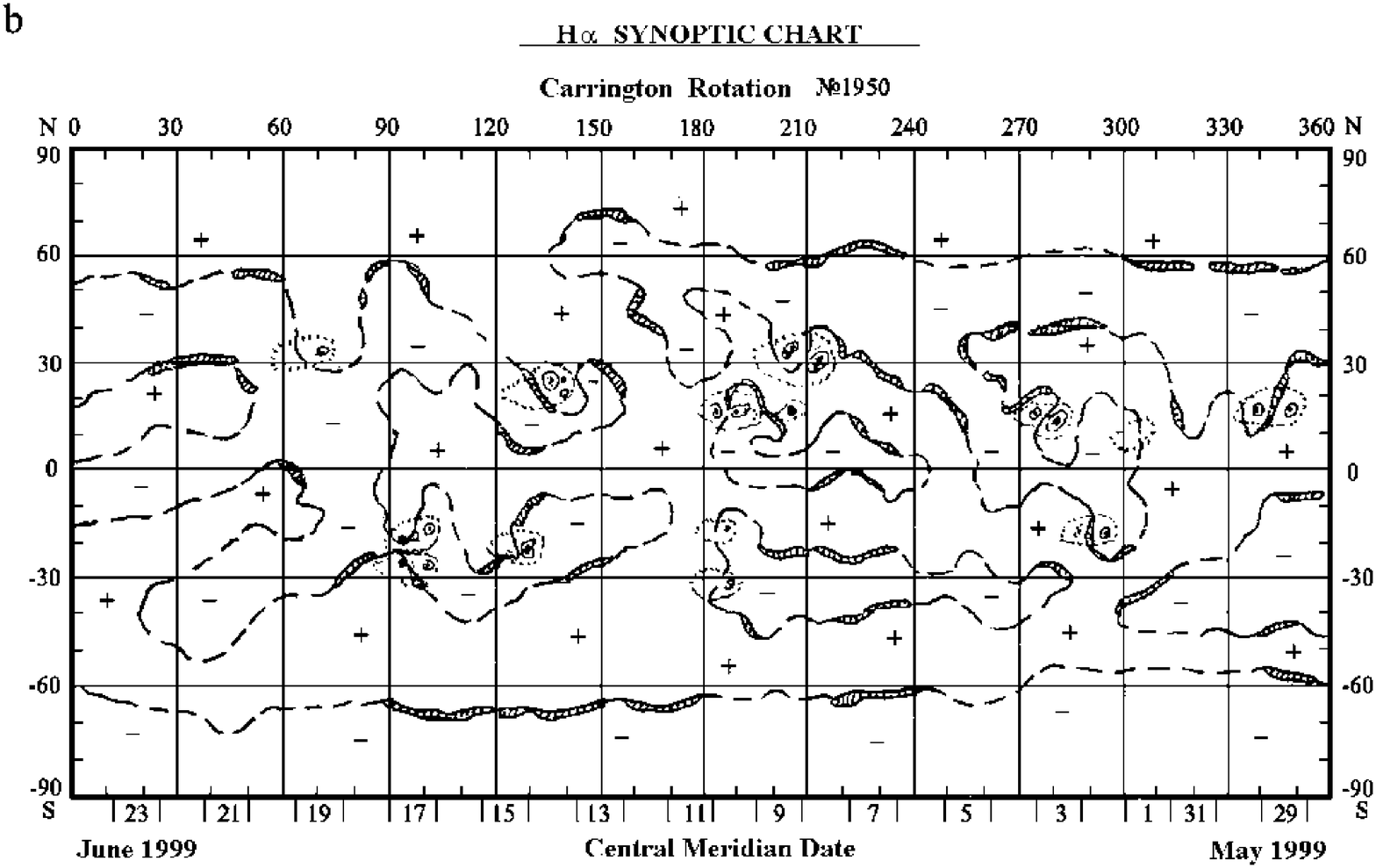}
    \caption{Meudon synoptic map for rotation N 1950 (a).
               Kislovodsk synoptic map for rotation N 1950 (b).}
    \label{Fig1}
\end{figure}

The filaments with $\tau<0$ in the mid-latitude region are mainly associated with active regions (see Figure~\ref{Fig1} and the scheme in (\cite{Mackay08}). Figure~\ref{Fig12}b represents the mean tilt angles for even and odd cycles. The tilt angles for even cycles are smaller than in the subsequent odd cycles which corresponds to the $20\%$ alternating variability in odd/even cycles according to Gnevyshev-Ohl rule (\cite{Gnevyshev}).

In the vicinity of the equator, the tilt angles are
highest in absolute value (Figure~\ref{Fig9}a). This suggests that the magnetic fields of opposite polarity alternate with longitude in accord with the concept of sector structure of the low-latitude large-scale magnetic field. At high latitudes, the tilt angles of filaments are small in accord with the concept of the zonal structure of the high-latitude field.

We used the data from $H_\alpha$ synoptic charts obtained at Kislovodsk to reconstruct the maps of the magnetic field polarities for the period from 1887 to the present day (Figure~\ref{Fig1}b) (\cite{Makarov89}; \cite{Makarov00}; \cite{Tlatov}).
We also compared the positions of filaments with the polarity of the large-scale magnetic field.
Figure~\ref{Fig13} shows variations of tilt angles for the filaments lying on the polarity separation line of the large-scale magnetic field ($+/-$)  where "+" corresponds to the
case when positive magnetic field is near to the north pole and "-" when the negative field is near to the south pole),
 and for those lying on the polarity separation ($-/+$). This is in agreement with the results of (\cite{Leroy}).  The filaments lying on ($+/-$) polarity separation have the maximum tilt angles in odd cycles, and the filaments lying on ($-/+$) polarity separation, in the even cycles. This may reflect the presence of the meridional component of the global magnetic field  i.e. the mean poloidal field.

\section{Conclusions}

Our paper deals with statistics of quantitative characteristics of solar filaments in the century-long period 1919-2014.
We especially focus on analysis of the tilt angles of filaments. We have shown that the tilts of the filaments of different kinds have different average values of tilts and have different regularities of long-term variability.
The mean tilt angles decrease in the epoch of minimum of the 11-year sunspot activity cycle  and have probably  positive correlation with the amplitude of the sunspot cycle (Figure~\ref{Fig7}).
Over the secular Gleissberg cycle, the tilt angles in the mid-latitude zone $\pm40^\circ$ were maximum in the middle of the 20th century in cycles 18-19.
We also found latitudinal migration of tilt filaments with the solar cycle both equatorwards in low latitudes, and polewards in high latitudes.

For the filaments which are associated with solar active regions and have negative tilts, the tilt angles are larger in odd cycles than in the preceding even cycles (Figure~\ref{Fig12}b).
For filaments lying within one the same system of the large-scale magnetic field polarity separation lines [either (+/-), or (-/+)], there is 22 year periodicity in variation of the tilts $\tau$ (Figure~\ref{Fig13}).
This may be connected with variation of latitudinal extent of the filaments which belong to different polarity systems (\cite{Makarov83}, \cite{Makarov89}), and the above described dependence of the tilt angles of filaments with their latitude.

\begin{figure}
  \centering
  \includegraphics[width=3.0in]{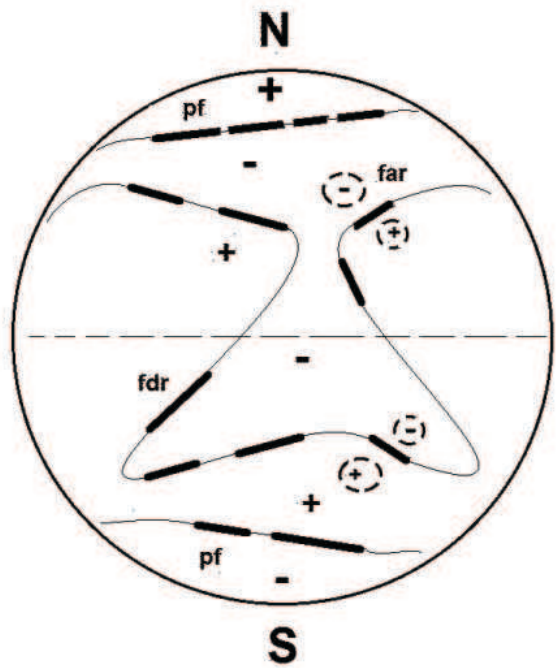}
  \caption{The scheme of classification of filaments with different tilt angles $\tau$. far - filaments of active regions ($\tau<0$), fdr - quiescent filaments between diffuse fields of different polarity  ($\tau > 0$), pf - polar filaments ($\tau<0$).  }
   \label{Fig2}
 \end{figure}

\begin{figure}   
   \centering
   \includegraphics[width=10.0cm]{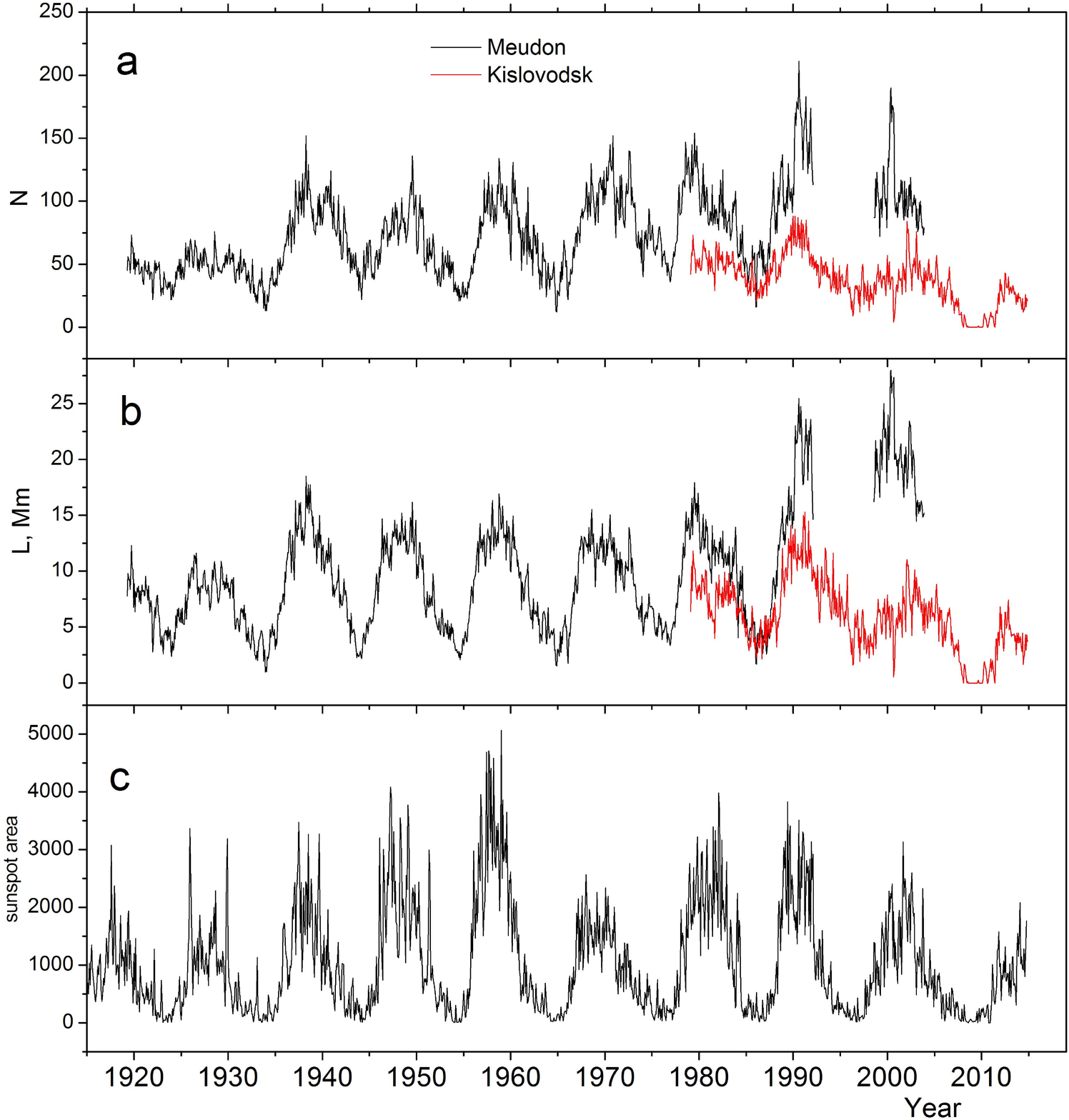}
   \caption{(a) Total number of filaments on synoptic charts; (b) total length of filaments on synoptic charts measured in Gm  ($10^9$ m);
               (c) monthly mean sunspot areas shown  for comparison in millionths of hemisphere (MHS)  for comparison.
 Cycle numbers are given below.
  }
 \label{Fig3}
  \end{figure}

\begin{figure} 
   \centering
\includegraphics[width=10.0cm]{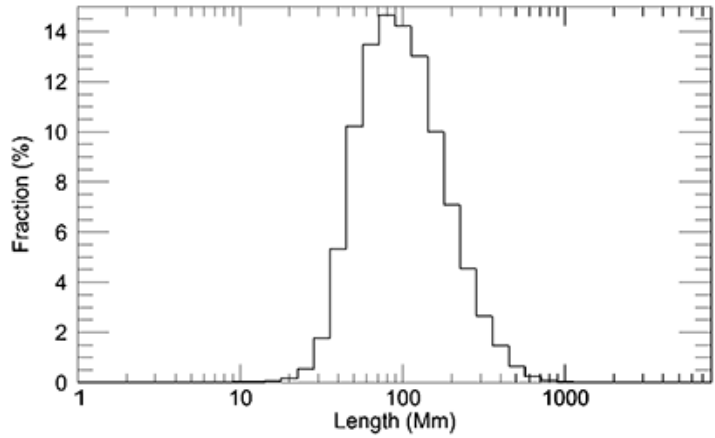}
   \caption{ Distribution of the filament length in logarithmic scale. The relative number of filaments according to their length in Mm.
        }
   \label{Fig4}
 \end{figure}

\begin{figure}    
    \centering
  \includegraphics[width=10.0cm]{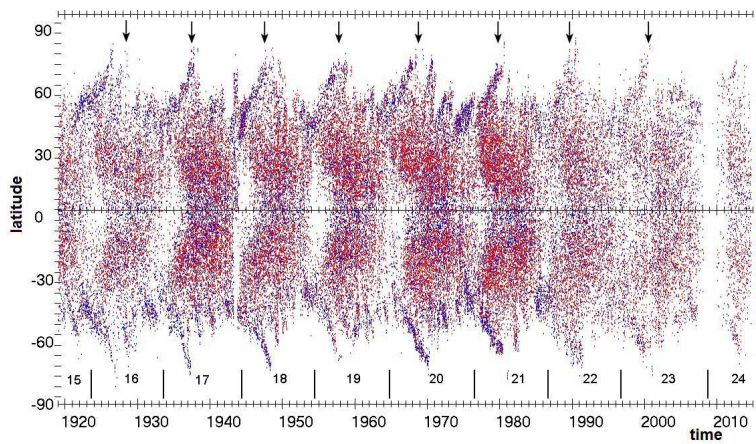}
   \caption{Latitude-time distribution of the  centers of filaments. Red dots correspond to the filaments with positive tilt angles $\tau$, and blue dots to the filaments with negative tilt angles.
  Arrows above indicate the times of maximum solar cycle while vertical bars below indicate the times of the minimum.
}
   \label{Fig5}
   \end{figure}

\begin{figure}    
    \centering
  \includegraphics[width=4.0in]{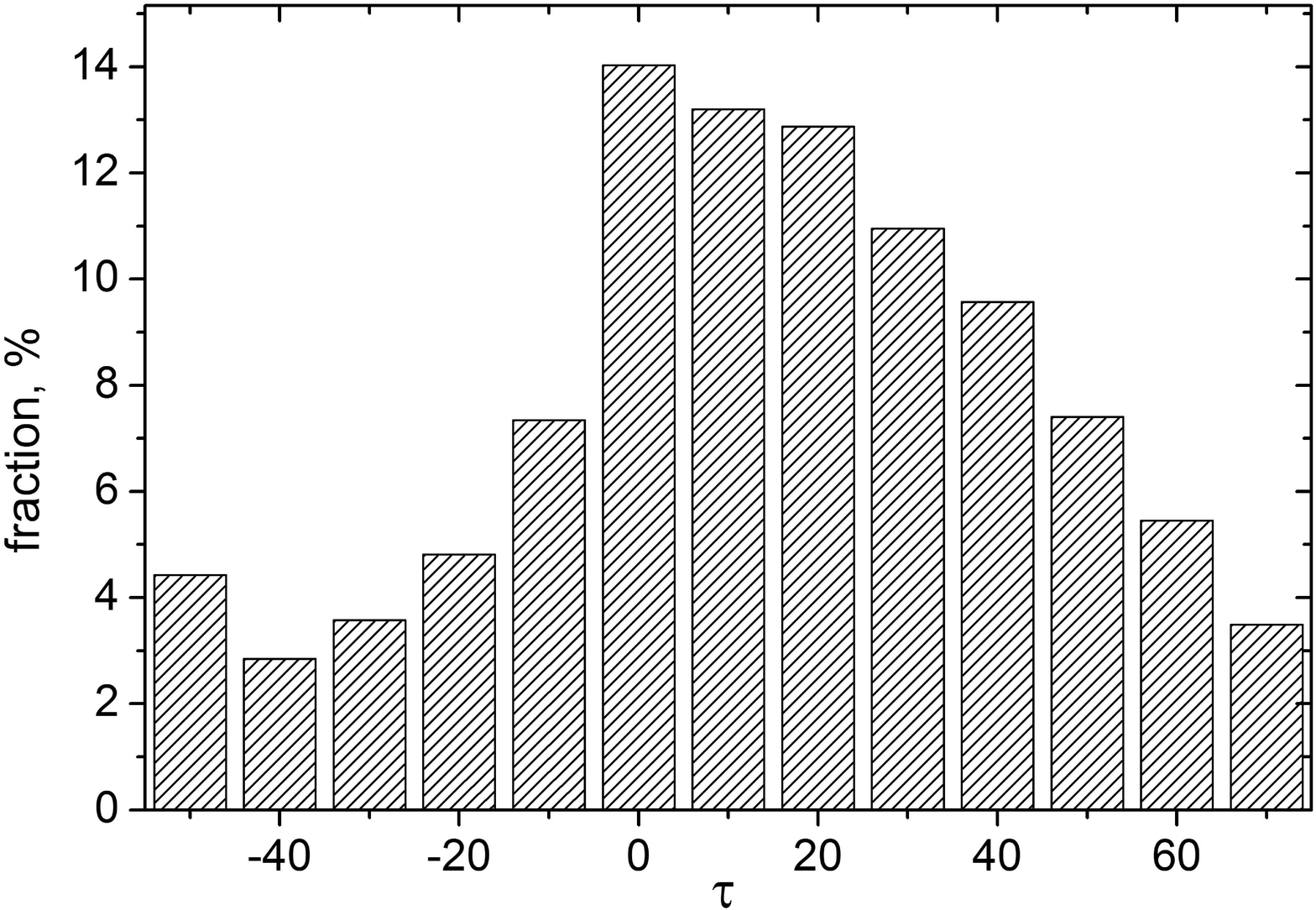}
   \caption{Distribution of the tilt angle of solar filaments with respect to the equator. The tilt angles of filaments in the southern hemisphere are taken with the opposite sign. $\tau=0^\circ$ corresponds to the filaments to be parallel to the equator; the positive angle means that the eastern tip of the filament is closer to the pole than the western one. }
   \label{Fig6}
   \end{figure}

\begin{figure}
   \centering
  \includegraphics[width=4.0in]{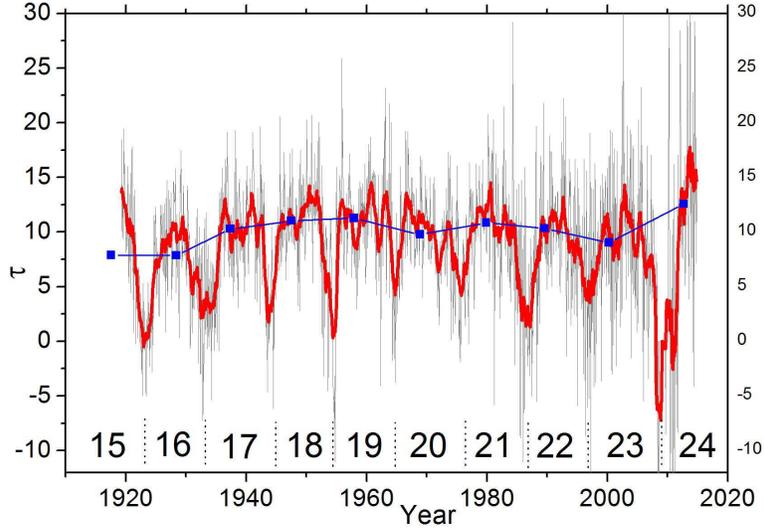}
              \caption{Variations in the filament tilt angle $\tau$ in degrees.
                 The data averaged over each synoptic rotation are shown as a weak grey line, and the same data smoothed by running average for 13 rotations are shown as a thick red line.
                 The mean tilt angles of all filaments in cycles 15-24 from $T^{n}_{min}$ to $T^{n+1}_{min}$ are shown by blue color; their Student's 90\% confidence intervals are of order one degree. Cycle numbers are given below.
                 }
   \label{Fig7}
   \end{figure}

\begin{figure}
   \centering
  \includegraphics[width=3in]{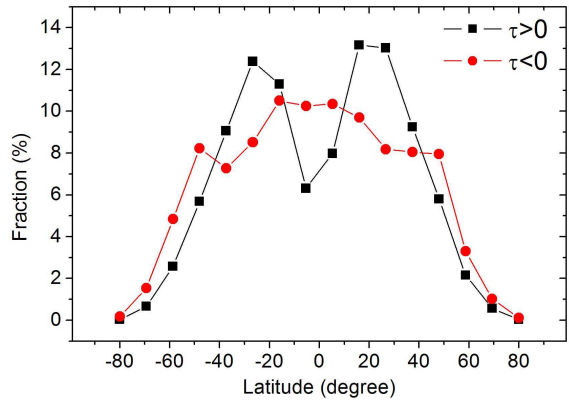}
              \caption{The relative number of filaments versus latitude for the filaments with the tilt angle $\tau>0$ (squares) and $\tau<0$ (circles). }
   \label{Fig8}
 \end{figure}

\begin{figure}
   \centering
  \includegraphics[width=4in]{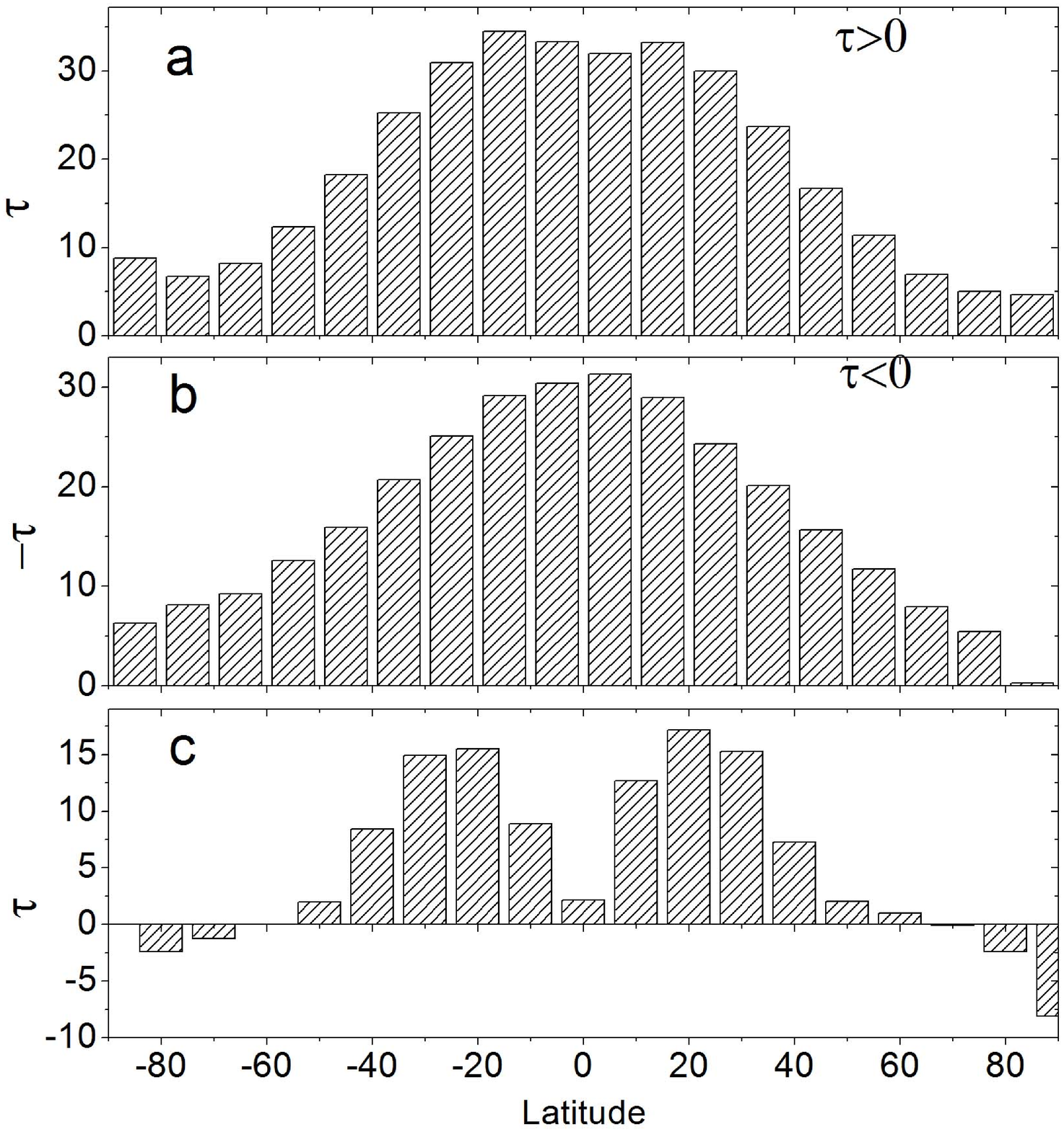}
              \caption{The tilt angles versus latitude (in degrees) for the filaments with positive (a), negative (b) tilts, and for all filaments (c). }
   \label{Fig9}
 \end{figure}

\begin{figure}
    \centering
   \includegraphics[width=5in]{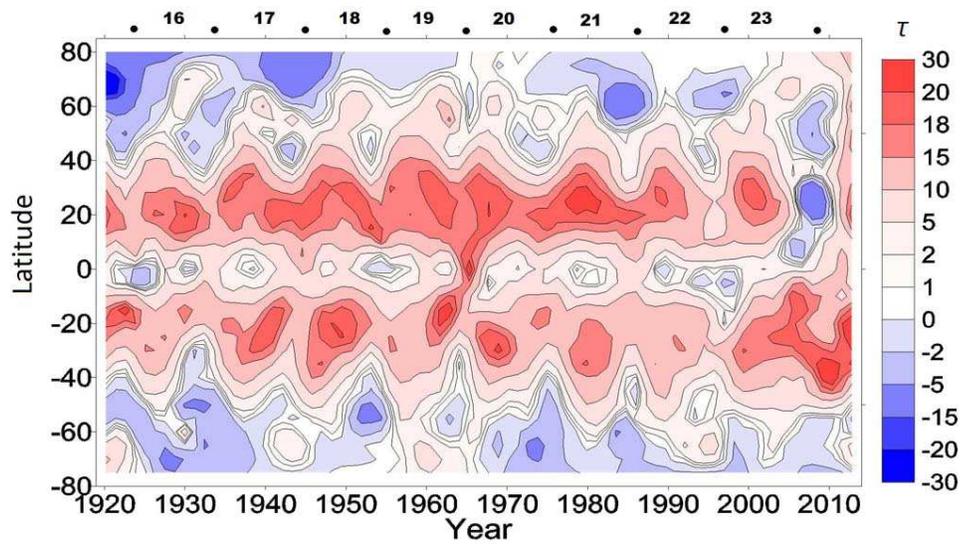}
              \caption{Distribution of the tilt angles of filaments. At mid-latitudes, the leading (western) tips of the filaments in the northern and southern hemispheres are closer to the equator than their tail (eastern) tips (the tilts are positive). Solar cycle numbers are shown above.}
   \label{Fig10}
 \end{figure}

\begin{figure}    
   \centering
     \includegraphics[width=4in]{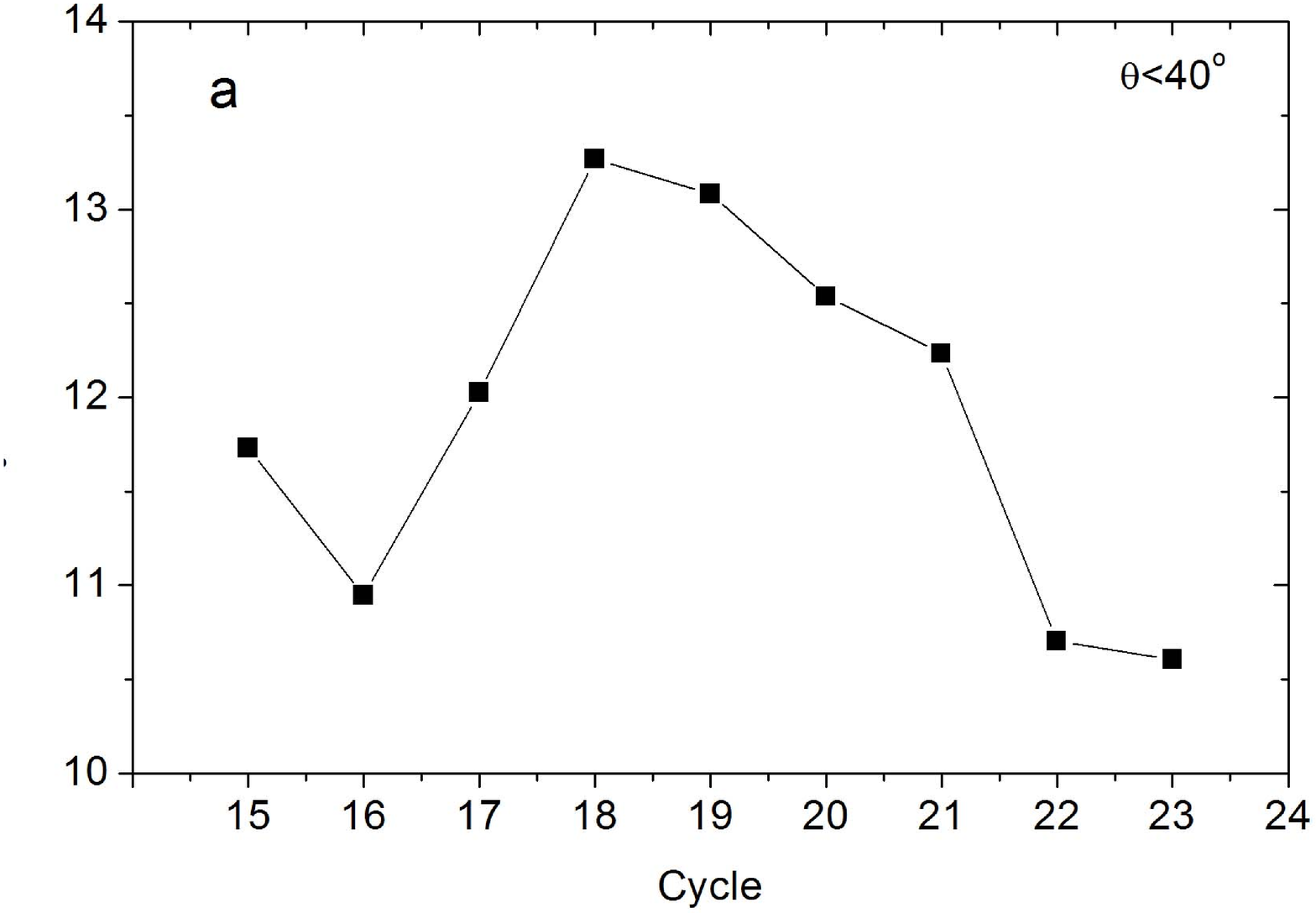}
    \includegraphics[width=4in]{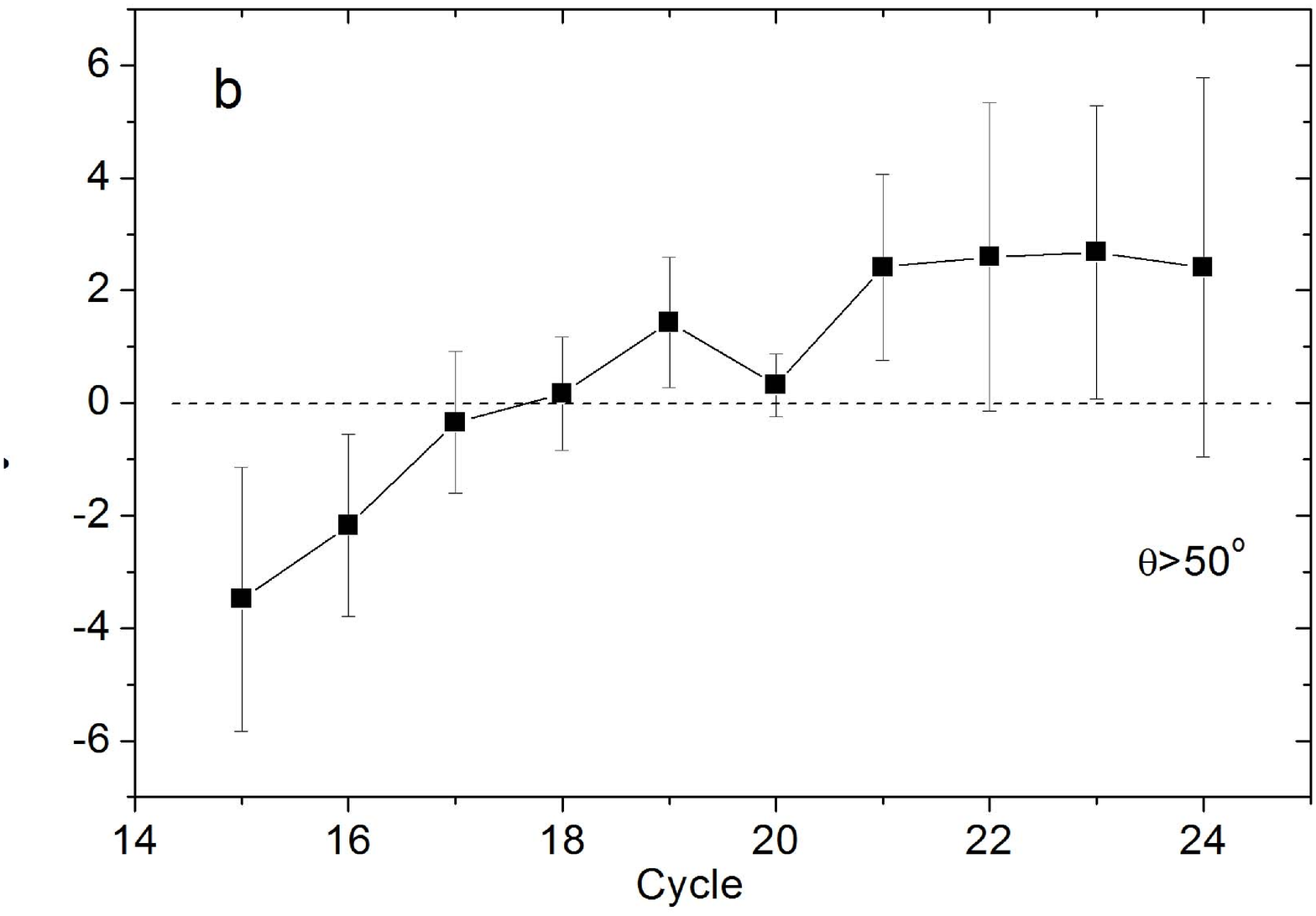}
\caption{Cycle variations in the tilt angles of equatorial (a) and polar (b) filaments. Error bars represent Student's $90\%$ confidence intervals. For filaments in equatorial zone error bars are indistinguishable (less than $1\%$ of the mean values).
        }
   \label{Fig11}
   \end{figure}

\begin{figure}    
  \centering                              
   \includegraphics[width=4in]{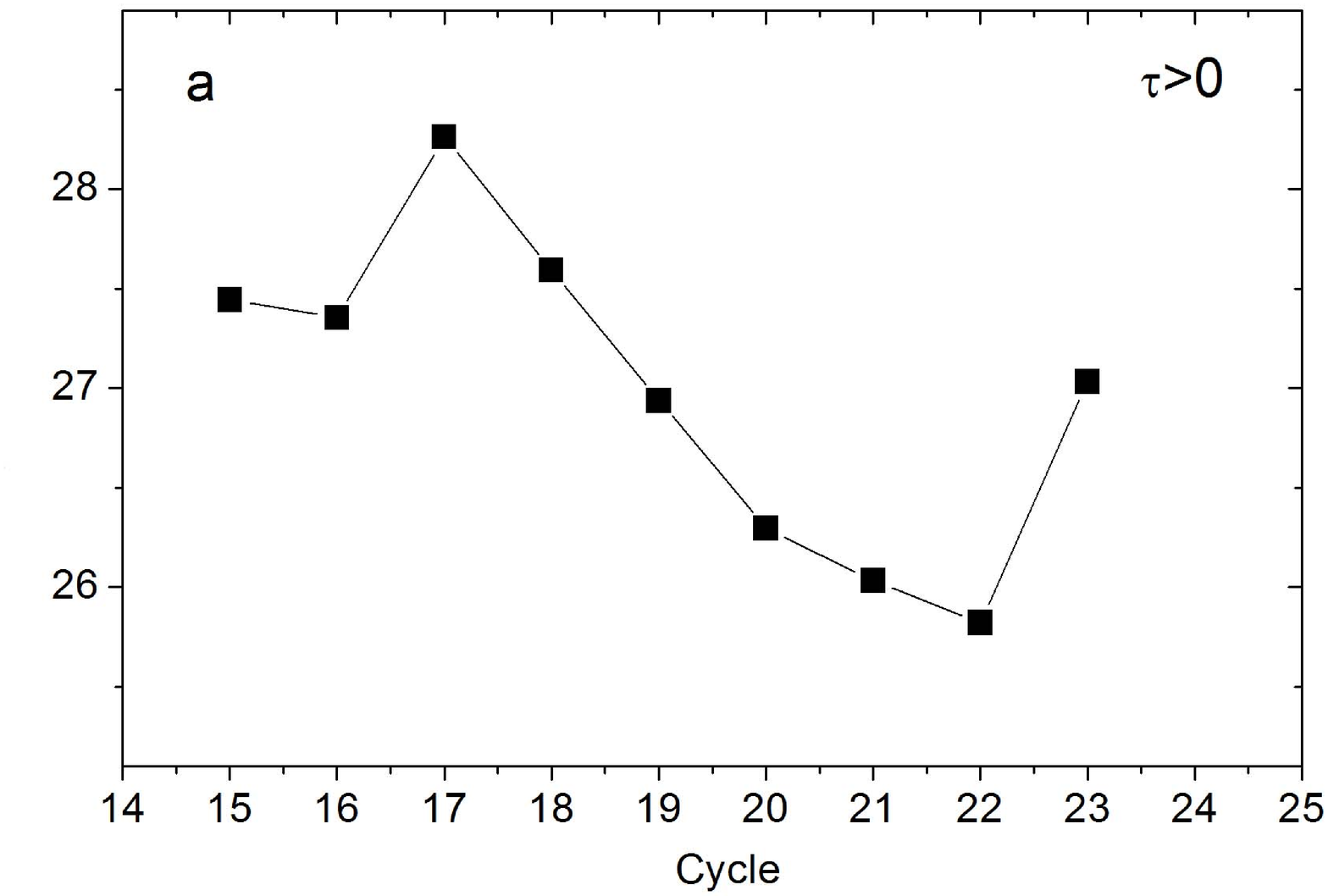}
   \includegraphics[width=4in]{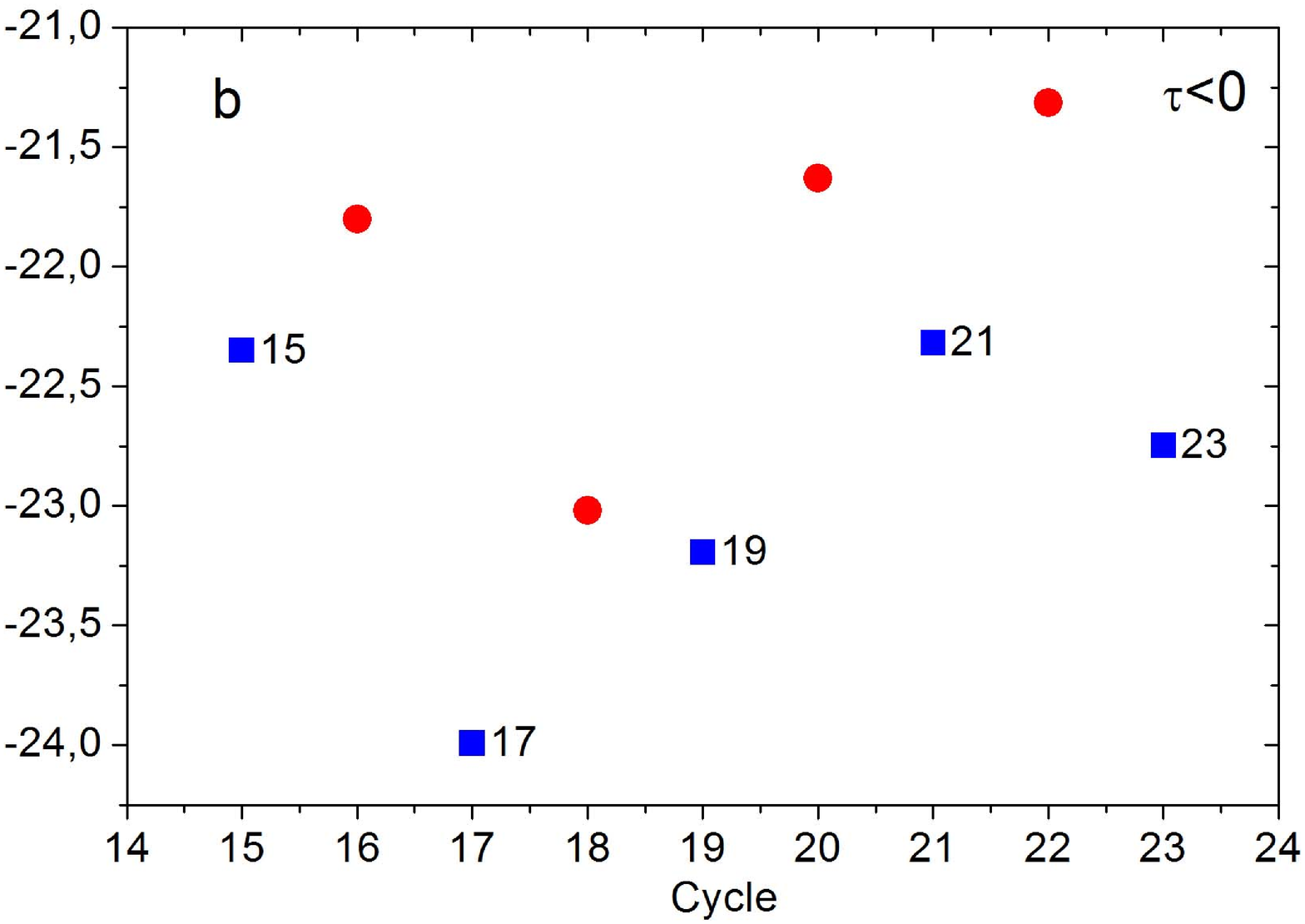}
\caption{ Cycle variations of the mean tilt angles of filaments with $\tau>0$ (a) and $\tau<0$ (b). The Student's $90\%$ confidence intervals error bars are indistinguishable (about $1-2\%$ of the mean values).
        }
   \label{Fig12}
   \end{figure}

\begin{figure}
    \centering
  \includegraphics[width=4in]{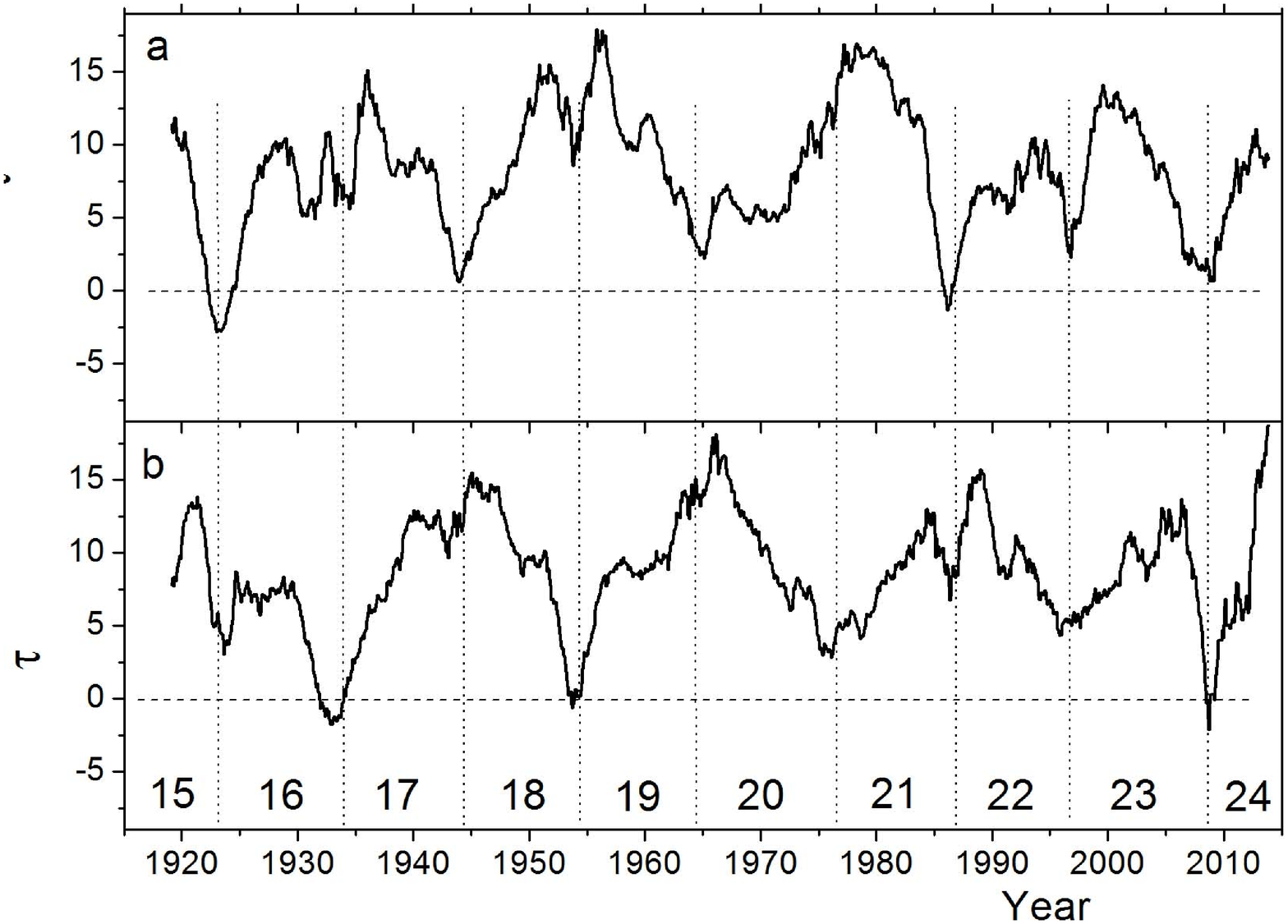}
  \caption{Tilt angles of filaments separating large-scale magnetic fields in latitude +/- (top) and in polarity -/+ (bottom). The data for each synoptic rotation are smoothed with running window of 27 Carrington rotation periods.
  Vertical dashed lines indicate solar cycle minima. Cycle numbers are given below.
               }
   \label{Fig13}
   \end{figure}

\vbox{
Therefore, we have established the three main classes of filaments with respect to their tilt angles
(see schematic diagram in Figure ~\ref{Fig2}):

\begin{itemize}

\item{}
Low latitude intra-active region filaments that have mainly negative tilts ("far");

\item{}
Low latitude extra-active region and/or extra-diffuse plage filaments that have mainly positive tilts ("fdr");

\item{}
High latitude extra-diffuse plage filaments that have mainly negative tilts  ("pf").

\end{itemize}
}

The statistical properties of solar filaments can be used as additional coherent measure of manifestation of the solar cycle which covers all latitudes.
The data series on filaments is systematically calibrated and is available for almost a century long period. This opens new perspectives for studies of the solar cycle and underlying hydro-magnetic dynamo mechanisms using this universal and long term systematically measured index of manifestations of the solar activity.

The Authors would like to thank the anonymous referee for useful and helpful criticism and suggestions for improving  the paper.
A.T. would like to acknowledge support from  Russian Science Foundation No. 15-12-20001,  K.K. would like to acknowledge partial support from the Russian Foundation for Basic Research under grant No. 13-02-01183, and  V.V. also partial support under grants RFBR No. 15-02-03900 and 15-02-10069.

\end{document}